\documentstyle[aps]{revtex}

\newtheorem{theorem}{Theorem}
\newtheorem{acknowledgement}[theorem]{Acknowledgement}

\begin{document}
\title{Photon counting for quantum key distribution with Peltier cooled InGaAs/InP
APD's. }
\author{Damien Stucki, Gr\'{e}goire Ribordy, Andr\'{e} Stefanov, Hugo Zbinden}
\address{{\small {\em Group of Applied Physics, University of Geneva,\\
1211 Geneva 4, Switzerland\newline}}
}
\author{John G. Rarity, Tom Wall}
\address{{\small {\em S\&E Division Defence Evaluation and Research Agency,\\
St Andrews Rd, Malvern, UK WR14 3PS \newline}}}
\date{\today }
\address{FINAL\ VERSION}
\maketitle

\begin{abstract}
The performance of three types of InGaAs/InP avalanche photodiodes is
investigated for photon counting at 1550 nm in the temperature range of
thermoelectric cooling. The best one yields a dark count probability of $%
2.8\cdot 10^{-5}$ per gate (2.4 ns) at a detection efficiency of 10\% and a
temperature of -60$%
{{}^\circ}%
$C. The afterpulse probability and the timing jitter are also studied. The
results obtained are compared with those of other papers and applied to the
simulation of a\ quantum key distribution system. An error rate of 10\%
would be obtained after 54 kilometers.
\end{abstract}

\section{Introduction}

Quantum Key Distribution (QKD), the most advanced technology of the field of
quantum information, allows two remote parties to exchange a sequence of
random bits and subsequently check their secrecy \cite{RMP}. It has been
extensively tested in the past couple of years over distances of a few tens
of kilometers \cite
{Townsend98,Bourennane99,Ribordy2000,Hughes2000,Bethune2000}. Its security
relies on the fact that the bits are encoded on single photons.

A first consequence of this fact is that the channel transporting the
photons, usually an optical fibre, must be as transparent as possible. The
attenuation in silica, which is fundamentally limited by Rayleigh
scattering, decreases with the wavelength. From around 2 dB/km at 800 nm, it
goes down to 0.35 dB/km at 1300 nm and even 0.2 dB/km at 1550 nm. Maximizing
the transmission requires to work at 1550 nm. At this wavelength,
respectively 10\% and 1\% of photons injected in a fibre are still present
after 50 and 100 km (as will be seen below, this value is slightly
overoptimistic).

Second, single photons are difficult to detect. Although commercial
detectors based on silicon avalanche photodiodes (APD) exist for wavelength
below 1000 nm, this is not true for wavelengths beyond. In order to detect
single photons at 1550 nm, commercial InGaAs/InP APD's designed for
telecommunication applications have to be operated in so-called Geiger mode.
In this mode, they are biased beyond the breakdown voltage. As soon as a
carrier is created within the diode, it sweeps through the junction and
triggers an avalanche which in turn yields a macroscopic current pulse. The
carrier can be photogenerated, in which case the current pulse gives
information about the arrival of a photon. It can also be generated by
spurious mechanisms -- like thermal excitation --\ which cause noise. After
the detection of the avalanche, it must be quenched in order to reset the
detector and prepare it for subsequent photons. Various techniques,
thoroughly discussed in \cite{Cova96}, can achieve this.

Single-photon detectors are a key component in a QKD system. They influence
both the key creation rate and the error rate. Although commercial
InGaAs/InP APD's have already been thoroughly tested in Geiger mode \cite
{Lacaita96,Ribordy98,Rarity2000,Hiskett2000}, we intend in this paper to
present recent results of the characterization of such APD's and interpret
them in the light of QKD. After a general introduction and the description
of the set-up, we present experimental results. We look at the detection
efficiency, the dark count probability, the afterpulse probability and the
timing jitter. Finally, we compare these values with those presented in
other works and use them in the simulation of a QKD system to estimate the
possible performance.

\section{Single-photon detection for quantum key distribution}

Most QKD systems are synchronous, in the sense that a timing signal can be
available to indicate the possible arrival of a photon. This fact makes it
possible to use the APD's in gated mode, where a biasing pulse is applied by
a suitable generator to gate the detector on. We only consider this mode
which is well suited to QKD.

An ideal single photon detector should produce an electronic logical signal
when and only when a photon strikes it. Real detectors unfortunately differ
from this simple picture. First, the detector sometime fails to record a
photon. The probability for an impinging photon to be detected -- also
called detection efficiency -- is lower than 100\%. Second, the detector
also has a non-zero probability to produce a count even though no photon is
present. Such an event can stem from the thermal generation of a carrier in
the sensitive area. In this case, it is known as a dark count. It can also
arise from the release of a charge trapped in the junction in the course of
a previous avalanche, in which case it is called an afterpulse. Finally, the
time between the absorption of a photon, the triggering of the avalanche and
its detection will be statistically distributed around an average value.
This uncertainty is called the timing jitter.

In order to work well for QKD, a single-photon detector should have a
reasonably high detection efficiency, and low dark count and afterpulse
probabilities. Although it is not very critical, there are also constraints
on the timing jitter. A given QKD system can be characterized by its raw bit
rate $R$ and its quantum bit error rate $QBER$ . The bit rate is basically
equal to the product of the probability $p_{T}$ of a photon to reach the
receiver's detector, the detection efficiency $\eta _{det}$, and the
repetition frequency $f_{rep}$.

\begin{equation}
R=p_{T}\cdot \eta _{det}\cdot f_{rep}
\end{equation}

The QBER is equal to the ratio of erroneous counts over total counts. We
will consider only errors caused by the detectors, which constitute the
dominant contribution. It can be written in terms of probabilities and is
essentially equal to the ratio of the probability to record a false count --
the sum of the dark count probability ($p_{dc}$) and afterpulse probability (%
$\left( p_{T}\cdot \eta _{det}+p_{dc}\right) \cdot \sum\limits_{n=1}^{\infty
}p_{ap}\left( \Delta t_{n}\right) $) -- over that of recording a correct
one. Note that $\Delta t_{n}=n\cdot \frac{1}{f_{rep}}$ and $p_{ap}\left(
\Delta t\right) $ is the probability to register an afterpulse during a gate
arriving $\Delta t$ after one that yielded an avalanche. In the low dark
count limit ($p_{dc}\ll 1)$ we can ignore afterpulsing from dark counts and

\begin{equation}
QBER=\frac{\text{incorrect counts}}{\text{total counts}}\approx \frac{%
p_{dc}+p_{T}\cdot \eta _{det}\cdot \sum\limits_{n=1}^{\infty }p_{ap}\left(
\Delta t_{n}\right) }{p_{T}\cdot \eta _{det}}=\frac{p_{dc}}{p_{T}\cdot \eta
_{det}}+\sum\limits_{n=1}^{\infty }p_{ap}\left( \Delta t_{n}\right)
\end{equation}

Without going too much into details (see \cite{RMP} for a discussion of
these quantities), a system should yield a $R$ as high as possible and a $%
QBER$ as low as possible. A low detection efficiency causes both a reduction
of $R$ and an increase of $QBER$. High dark count and afterpulse
probabilities yield high $QBER$. The dark count probability can nevertheless
be reduced by cooling the detector. Even though cryogenic temperatures are
easy to obtain in the laboratory, for example by using liquid nitrogen, they
are not suited to the realization of a commercial prototype. We only
consider here the temperature range that is easily attainable with
thermoelectric cooling (down to $-60%
{{}^\circ}%
C$). The dark count probability can also be reduced by recording counts in a
small time window around the arrival of the photons. The width of this
window is ultimately limited by the timing jitter.

\section{Detection efficiency and dark counts}

Three different types of InGaAs/InP APD's (Epitaxx EPM 239 AA, EG\&G 30733,
and NEC NDL 5551) were tested and compared. Although the results presented
in this paper were obtained with one single Epitaxx APD, additional tests
performed on other specimens of this APD and not reported here yielded
similar results in photon counting mode. As for the EG\&G and NEC APD's,
only one APD of each was tested. Thus we cannot present information about
differences among samples. Both the Epitaxx and the NEC APD's were mounted
on a copper block equipped with a heating coil. It was placed in a closed
tube and immersed in liquid nitrogen. The temperature was then adjusted by
heating the arrangement. This solution was selected because of the
availability of the equipment, but a Peltier cooler would have been just as
suitable for the target temperature range. The EG\&G APD is mounted in a DIL
package and includes a Peltier cooler and thermistor. The whole package was
then mounted on a single-stage peltier cooler in an insulated metallic box.
The hot side of the thermoelectric element was placed on a radiator equipped
with a fan.

The schematic set-up for measuring the quantum detection efficiency and the
dark count probability of the APD's is shown in Fig. 1. A delay generator
(Stanford Research Systems DG 535) acts as the time base of the system and
generates signals, unless otherwise mentioned, at a frequency of 10 kHz. It
triggers first a pigtailed semiconductor laser (MRV MRLD14CD5015), which
generates short light pulses ($\simeq $ 350 ps) at 1550 nm. These pulses are
attenuated to the single photon level by a calibrated variable attenuator
(EXFO IQ-3100, maximum attenuation 
\mbox{$>$}%
100 dB) equipped with FC/APC connectors. The output of this attenuator is
connected to the pigtail of the APD. The attenuator has a shutter that can
be used to turn on and off the illumination of the detector.

The delay generator also triggers a gate generator that produces square
biasing pulses with an amplitude of 6.5 V and a full width at half maximum
of 2.4 ns (see Fig. 2). The delays are set so that the photons impinge on
the APD's sensitive area when it is gated on. This pulse is superimposed
upon an offset voltage, whose level is adjusted to control the excess bias
voltage. The avalanche signal is then detected with a discriminator which
generates logical output pulses. These are finally registered with a
counter. As can be seen in Fig. 2, the rising and falling edges of the gate
pulse are sharp. They produce transients in the output signal that can be
larger than the avalanche itself, especially for low excess bias voltages.
The output of the discriminator is fed into a coincidence circuit in order
to reject the transients and record only the avalanche signals.

For a given bias voltage, we measured first the dark count probability. The
shutter was closed, and counts accumulated. We selected the integration
time, to keep the statistical uncertainty typically around 5\%, except for
the lowest bias voltage. In this case, the low dark count probability
prevented us from going below 10\% (accumulation of at least 100 counts).
The dark count probability is then simply calculated as the ratio of the
number of counts over the product of the repetition frequency and
integration time. In order to obtain the detection efficiency, we opened the
shutter and registered counts for a given time. It was then estimated by
subtracting the dark count probability. The Poissonian statistics of the
number of photons per pulse was taken into account. We then changed the
excess bias voltage and repeated the measurements.

The results for the three APD's are shown in Fig. 3, where the dark count
probability is plotted as a function of the detection efficiency. The figure
shows the error bars stemming from statistical uncertainties in the case of
the Epitaxx-APD for the detection efficiency. Those concerning the dark
count probability are not shown since they are smaller than the symbols. For
the other APD's, the uncertainties are similar but were omitted for the sake
of clarity. Considering first the data corresponding to the Epitaxx APD at
-40$%
{{}^\circ}%
$C and -60$%
{{}^\circ}%
$C, we notice a positive correlation between these two quantities. The
points correspond to different excess bias voltage. Both the detection
efficiency and the dark count probability increase with excess bias voltage.
This reflects the fact that a strong applied field enhances the probability
for a carrier to start an avalanche by impact ionization. The data can be
well approximated by an exponential fit. The slope is identical for both -40$%
{{}^\circ}%
C$ and -60$%
{{}^\circ}%
C$. The results also show that the dark count probability can be reduced by
cooling the detector. When the temperature is reduced, the curves shift
downwards. At a detection efficiency of approximately 10\%, the dark count
probability is equal to $6\cdot 10^{-5}$ and $2.8\cdot 10^{-5}$ at -40$%
{{}^\circ}%
C$ and -60$%
{{}^\circ}%
C$ respectively.

We verified that the contribution of afterpulses to the measured data is
negligible. The repetition frequency is indeed so low that the traps filled
by an avalanche have already been emptied when the next gate pulse is
applied. To establish this, we changed the repetition frequency and verified
that the dark count probability did not change. At the maximum repetition
frequency of 800 kHz and T = -40$%
{{}^\circ}%
C$, we did not measure any increase. In the next, section we will discuss
the issue of afterpulses in more detail.

The NEC-APD and the EG\&G-APD show similar performance (see also Fig. 3),
but are not as good as the Epitaxx-APD. Their dark count probability is
around $3.5\cdot 10^{-4}$ at an efficiency of 10\% (T = -60$%
{{}^\circ}%
C$), while it is as low as $2.8\cdot 10^{-5}$, more than one order of
magnitude smaller, for the Epitaxx-APD in similar conditions.

It is clear that for applications where the dark count probability is an
important issue, cooling the detector has a significant impact. Since it is
critical for QKD, one could wonder why we did not test the detectors to
temperatures lower than -60$%
{{}^\circ}%
C$. The first reason is that we want to be able to cool the detectors with
thermoelectric elements (Peltier coolers). They are indeed practical and
reliable. One could nevertheless object that other proven technologies (e.g.
Stirling cycle coolers) that allow to reach lower temperatures exist. This
would however not help us. More importantly, when the temperature is
lowered, the breakdown voltage $V_{B}$ decreases. As explained by Cova \cite
{Lacaita96} as well as Hiskett \cite{Hiskett2000}, $V_{B}$ should always
remain larger than the so called reach-through voltage $V_{RT}$ if one wants
to have the carrier generated in the absorption layer injected in the
high-field multiplication zone. As $V_{RT}$ does not change with
temperature, contrary to $V_{B}$, the APD might breakdown at a voltage
smaller than the reachthrough level below a certain temperature. The
Epitaxx-APD for example did not operate below -90$%
{{}^\circ}%
$C. Finally, when cooling the detectors, the lifetime of the trapped
charges, that cause afterpulses, increases.

\section{Afterpulses}

We investigate now the afterpulse probability at different temperatures and
for the various APD's. The set-up is similar to the one used in the previous
section (see Fig. 1). A third timing signal provided by the delay generator
is combined through an OR-gate to the one sent to the gate generator. The
APD is thus turned on at two subsequent times separated by a variable delay $%
\Delta t$. The laser pulse is still synchronized with the first gate. When
disconnecting the OR-gate input corresponding to the second gate pulse, one
is back in the situation presented in the previous section. This arrangement
was used to measure the detection efficiency and dark count probability, and
verify that the APD was operating properly. The results presented below were
all obtained with a bias voltage corresponding to a detection efficiency of
approximately 10\%. With both input of the OR-gate connected, one could
count events happening both in the first and second gate pulse. In order to
register only the events where a count happened both in the first and the
second window, we used a coincidence circuit. One can then use these data to
evaluate the probability to register a pulse in the second gate knowing that
one was obtained in the first one for various time delays between both
gates. After subtracting the dark count probability, one obtains the
afterpulse probability $p_{ap}\left( \Delta t\right) $.

The results for the Epitaxx-APD are shown in Fig. 4. The probability $p_{ap}$
is plotted versus the time interval $\Delta t$\ between the two gates, with $%
\Delta t$ up to 10 $\mu $s. It starts at approximately $10^{-2}$ and
decreases. One would expect that it would go down to zero for a time long
enough. The uncertainty nevertheless increases with $\Delta t$, since the
number of counts accumulated becomes smaller. In addition, one expects $%
p_{ap}$ to become larger, for a given value of $\Delta t$, when the
temperature is reduced. The lifetime of the traps should indeed become
longer. While this seems to be true for times beyond 5 $\mu s$, it is not
before. Cova deduced from measurements with Si-APD's that several trapping
levels, with different time constants, contribute to afterpulses \cite
{cova91}. This is likely to remain true with InGaAs/InP APD's. A different
temperature dependence of the lifetime for the trapping levels might explain
this result. Figure 4 also suggest that temperature may not have a strong
impact on the afterpulse probability in the Peltier cooling range.

When considering the other APD's, one finds that their afterpulse
probability is both lower than that of the Epitaxx-APD and decreases faster
(Fig. 5). Differences among the different APD's is not surprising, since the
trap density depends intrinsically on the structural quality of a diode. It
is reasonable to conjecture that improvements could be obtained with
appropriate efforts by manufacturers.

These results also demonstrate that the afterpulse probability after 100 $%
\mu $s is very low (not shown on Fig. 4). When using a repetition frequency
of 10 kHz, the trapped charges are always released between two gates. This
justifies our claim that we were able to isolate dark counts effects from
afterpulses in the measurements presented in the previous section.

\subsection{Avalanche pulse analysis}

As explained by Cova and his coworkers in \cite{Lacaita96} and \cite
{Lacaita94}, it is possible to obtain valuable information about the
trapping phenomena in an APD by analyzing the avalanche pulse. To do that,
we had to modify slightly the experimental set-up. It is indeed necessary to
apply long gate pulses ($t_{Gate}\geq 20$ ns). Since the width of the gate
produced by our generator cannot be increased, we directly fed the variable
output of the delay generator into a custom made amplifier (maximum
amplitude 8V). We then recorded the avalanche pulses with a digital
oscilloscope (Tektronix TDS 580C, 1 GHz bandwidth). The APD's output could
also be fed into a time-to-digital (TDC) converter to record counts
occurring within a precisely defined time window. We used this possibility
to accumulate counts over a 2 ns long region of the gates, yielding results
that can be compared with those obtained with short gates. The rest of the
set-up was identical.

When looking at a typical trace (Fig. 6), one notices the avalanche pulse
and the transient corresponding to the falling edge of the gate pulse. The
avalanche increases sharply and reaches a maximum after 2.5 ns, before
reaching a regime where it remains constant for about 5 ns. Then, it
decreases to a plateau that lasts until the end of the gate. This decrease
indicates a change in the breakdown voltage of the junction during the
avalanche. They suggest two mechanisms for this effect. First, the fact that
traps get filled during an avalanche pulse changes the charge distribution
in the junction and increases $V_{B}$. The typical time constant of this
phenomenon is of the order of a few nanosecond. The second effect is that
the current flowing in the junction raises its temperature, which also
increases $V_{B}$. In this case the time constant is larger, typically a few
hundred nanoseconds. In order to observe it, even longer gate pulses are
necessary (results not shown here). Both of these phenomenon yield in turn a
reduction of the excess bias applied on the junction, and a decrease of the
pulse amplitude. Information about the diode structure (dopant
concentration), as well as the peak to plateau difference can be used to
estimate the fraction of filled traps. However this structural information
was not available. Nevertheless we can conclude from this avalanche pulse
analysis that the trapping effects are significant in the Epitaxx-APD's. In
addition, their filling takes about 5 -- 7 nanoseconds at a gate voltage
corresponding to an efficiency of about 10\%. This indicates that it is
essential for the duration of the gate pulse to be smaller than this time,
if one wants to reduce the afterpulse probability by a limitation of the
charge transit in the junction. When measuring quantitatively the afterpulse
probability with long gates (Fig. 7), we indeed observed results
approximately two orders of magnitude higher than those obtained with short
gates. After 100 ns, the afterpulse probability is close to 100\% when long
gates are used, while it is only of the order of 1\% with short gates. The
fact that avalanches are not quenched before the end of the gate, lasting 20
ns, explains this result. Note that in the case of long gates, the
afterpulses are accumulated in a 2 ns window defined with the TDC and not
over the entire gate duration. This guarantees that the results obtained
with the two types of gates can be compared.

\subsection{Gate amplitude}

Contrary to the short gate generator, the long gate generator allows to
change the gate amplitude and to investigate its effect on the performance
of the detector. Cova and his coworkers suggest that holding the detector
biased close to the breakdown voltage during the off-periods enhances
detrapping through the Franz-Keldish effect (see \cite{Lacaita96}). To
verify this, we used gate amplitudes of 3, 4, 6 and 8 V. In each case, we
made sure that the excess bias voltage was such that the detection
efficiency was about 10\% and measured the afterpulse probability as
discussed above. The results are shown in Fig. 7. It appears clearly that,
contrary to what we expected, a higher gate amplitude, which is corollary to
holding the detector well below the breakdown voltage during the
off-periods, reduces the time constant of the afterpulse probability. It
indicates that, in III-V hetero-junctions such as those of InGaAs/InP APD's,
other phenomena dominate the Franz-Keldish effect. Further research is
necessary to better understand them.

We also measured the dark count probability as a function of the quantum
detection efficiency for the various gates amplitudes with a repetition
frequency of 10 kHz (Fig. 8). The temperature was -50$%
{{}^\circ}%
C$. The results clearly indicate that the performance of the detectors is
identical in all cases. The amplitude of the gate influences thus only the
afterpulse probability. With long gates (20 ns) nevertheless a dark count
probability of $10^{-4}$ for a detection efficiency of 10\% was measured.
This performance is not as good as that obtained with short gates. This can
be explained by the higher afterpulse probability. In this case, the
repetition frequency was not low enough to allow the traps to empty
themselves.

In summary, if one wants to reduce the afterpulse probability when working
with InGaAs APD's, the gates should be kept as short as possible and their
amplitude as high as possible.

\section{Timing jitter}

The timing jitter is the last property of the APD's that we investigated. We
used the same set-up as above with long gates and fed the output of the
detectors into the TDC. When sending light pulses in coincidence with the
gate, we obtained time spectra consisting of a broad pedestal corresponding
to dark counts and a well defined peak corresponding to the light pulse. We
can deduce from the measured full width at half maximum (FWHM) of the peak
and that of the laser pulse ($\simeq $ 350 ps) the timing jitter of the APD.
Figure 9 shows the results obtained at three different temperatures and for
the Epitaxx-APD as a function of the detection efficiency. One can see that
the values typically range between 500 and 300 ps. The jitter decreases with
the efficiency. At 10\% its value is around 450. Finally, the temperature,
in the studied range, does not have a significant impact.

\section{Comparison with previous results}

Let us now compare the results obtained for the Epitaxx-APD with those
presented by Hiskett and his coworkers \cite{Hiskett2000} for the detection
of photons at 1550nm. They tested two different types of Fujitsu-APD's
(30FPD13U81SR with 80$\mu $m active area diameter, and FPD13W31RT with a 30$%
\mu $m active area diameter). When cooled to -196$%
{{}^\circ}%
$C, the first device yields a dark count probability of $10^{-4}$ in a 2.5
ns window for a detection efficiency of 10\%. As a reminder, we obtained for
the Epitaxx-APD a dark count probability of $2.8\cdot 10^{-5}$ at -60$%
{{}^\circ}%
$C and $\eta _{det} =10\%$. In spite of the fact that they do not measure
the afterpulse probability, they observe that for their APD cooled to -196$%
{{}^\circ}%
$C, the dark count probability already doubles when the repetition frequency
is increased to 30 kHz. Both the higher dark count and afterpulse
probability would yield lower performance in a QKD system with respect to
those that would be obtained with an Epitaxx-APD.

In addition, the APD's tested in \cite{Ribordy98} (Fujitsu FPD5W1KS) yield
similar performance as those presented by Hiskett ($P_{dc}=10^{-4}$ at $\eta
_{det} =10\%$). These results are however obtained at a temperature of -80$%
{{}^\circ}%
$C, where the APD's performance is optimal.

Finally, the EG\&G and NEC APD's had been previously tested using a passive
quenching approach\cite{Rarity2000}. They showed a higher dark count
probability ($p_{dc}\approx 2.5\cdot 10^{-3}$ per 2.4 ns with $\eta _{det}
=10\%$). A significant number of dark counts arose due to afterpulsing from
dark counts alone. This highlights the advantage of gated operation where
the only significant afterpulsing contribution comes from the light counts.
Afterpulsing was measured using a correlation technique and integrated over
long times so that it cannot be directly compared but appears to be of a
similar magnitude. When correcting the dark count probability for
afterpulsing, a value only slightly higher than in gated mode ($\geq 5\cdot
10^{-4}$) was found.

\section{Application to QKD}

Let us now use the results obtained here, to simulate QKD. We will first
neglect the effect of afterpulses. We will also consider that the detectors
are the only source of errors. The raw key creation rate can be calculated
using Equation (1). It is necessary however to consider $p_{T}$, the
probability for a photon to reach the receiver's detectors, in more details.
Let us rewrite this quantity as

\begin{equation}
p_{T}=\mu \cdot T_{L}\cdot T_{R}\text{.}
\end{equation}

In this equation, $\mu $ represents the probability for an emitted pulse to
contain at least one photon. Since ideal single photon sources do not exist,
one approximates them with attenuated laser pulses containing an average
number of photons of typically 0.1. Here, we will follow this convention and
set $\mu =0.1$. The second quantity, $T_{L}$, represents the transmission
probability of the a photon through the optical fibre line between the
emitter and the receiver. We will assume that the attenuation at 1550nm in
optical fibres is 0.25dB/km, somehow higher than the lowest attenuation
measured. Installed fibres tend indeed to show higher attenuation than the
absolute minimum and feature splices, typically every few kilometers.
Finally $T_{R}$ represents the transmission of the receiver system. We will
set it to 0.5, corresponding to an attenuation of 3dB.

Setting $\eta _{det}=10\%$, as we experimentally obtained, one can calculate 
$R/f_{rep}$ as a function of the distance in kilometers (Fig. 10). Even when
the distance is equal to zero, the normalized raw rate is smaller than
unity, because most pulses are empty, the detection efficiency is low and
the transmission of the receiver not perfect. The normalized rate then
decreases with the distance. Note that the sifted key rate, after basis
reconciliation, is half the raw rate plotted on Fig. 10.

We also plotted on Fig. 10 the $QBER$ calculated with Equ. 2 for the
Epitaxx-APD at -40$%
{{}^\circ}%
$C and -60$%
{{}^\circ}%
$C, for the EG\&G or NEC APD's at -60$%
{{}^\circ}%
$C, and finally for a hypothetical improved detector featuring a dark count
probability ten times smaller than that of the Epitaxx-APD. The impact of
this quantity appears clearly. Although the Epitaxx-APD yields satisfactory
results ($QBER=10\%$ after 54 km), the EG\&G or NEC\ APD's are not suitable
for QKD. Finally, the hypothetical detector would allow QKD over distances
well beyond 50 km. Note that the fraction of bits lost during key
distillation (error correction and privacy amplification) can be estimated
to 50\% and 85\% at a $QBER$ of 5\% and 10\% respectively. The error rate
should thus clearly be kept below 10\%.

These results were obtained assuming that afterpulses could be neglected.
This is true as long as the repetition frequency is slow enough. Looking at
Fig. 10 again, it appears clearly that the normalized raw rate decreases
steeply with distance. At 70 km for example, it is only $10^{-4}$ of the
repetition rate. In order to generate a non-negligible bit stream, a high
repetition frequency is necessary. The question now is to evaluate the
impact of afterpulses under these conditions.

Let us assume that we will tolerate an increase of the $QBER$ of 1\%, which
implies that $\sum\limits_{n=1}^{\infty }p_{ap}\left( \Delta t_{n}=n\cdot
1/f_{rep}\right) $ must be smaller than $10^{-2}$. Whether this condition is
fulfilled depends on the time dependence of $p_{ap}$ and the repetition
frequency. When considering the Epitaxx-APD with short gates at -60$%
{{}^\circ}%
$C and a repetition frequency of for example 1 MHz, one finds that the
cumulated afterpulse probability is equal to 1.4\%. Note that the values $%
p_{ap}\left( \Delta t\right) $ were obtained from an exponential decay fit
of the data shown on Fig. 4 and consisting of three terms. This probability
can be reduced by suppressing an appropriate number of gates after each
recorded detection, in order to wait for the trapping levels to empty. In
this example, such a hold-off mechanism must skip two gates after each
detection to bring the cumulated probability below 1\%, while the key
creation rate would be affected in a negligible way. For higher repetition
frequencies, the number of skipped gates must be increased (14 gates at 2
MHz). Finally, this analysis also indicates that long gates are not suitable
for QKD at a high repetition frequency because of the higher afterpulse
probability they induce.

Let us finally discuss the timing jitter of the detectors in the context of
QKD. This quantity is important for two reasons. First, it is possible to
reduce dark counts by time discrimination. Second, certain QKD set-ups
involve interferometers where different optical paths must be distinguished
in the time domain. These must clearly induce time differences in the
arrival of the photons larger than the jitter. Since interferometers with
large path differences are in general difficult to stabilize, a small jitter
constitutes an advantage. In order to keep the overlap of two gaussians
smaller than 5\% with a jitter of 450 ps, one must select a time difference
of at least 2.6 ns, corresponding in optical fibres to a length of 0.52 m
approximately.

\section{Conclusion}

In this paper, we have tested three types of InGaAs/InP APD's in the
temperature range of Peltier cooling at 1550 nm, in the light of QKD. We
have first investigated their detection efficiency and dark count
probability. The best APD -- the Epitaxx-APD\ -- yielded dark count
probabilities of $2.8\cdot 10^{-5}$ and $6\cdot 10^{-5}$ for temperatures of
-60$%
{{}^\circ}%
$C and -40$%
{{}^\circ}%
$C respectively and an efficiency of 10\%. We also measured the afterpulse
probability. The importance of limiting the duration of the avalanche -- and
use short gates -- to reduce this quantity was clearly demonstrated. The
timing jitter of the Epitaxx-APD was also investigated.

We compared the results obtained with those published in other papers on
InGaAs/InP APD's. The performance of the Epitaxx-APD constitutes the best
one reported to date for photon counting at 1550 nm. In addition, it is
obtained at a temperature easily reachable with Peltier cooling, whereas
other diodes required more sophisticated cooling.

These results were then applied to the simulation of a QKD system,
illustrating that among the three types of APD's, the Epitaxx-APD is the
only one to yield satisfactory performance. Error rates of 5\% and 10\%
would be obtained after respectively 40 km and 54 km. Assuming a repetition
frequency of 1 MHz, the raw key creation rate would be in the 100 Hz range.
Finally, they indicate the necessity to use a hold-off mechanism when
operating these APD's at a high repetition frequency.

\begin{acknowledgement}
The authors would like to thank Prof. Sergio Cova for helpful discussions.
This work was supported by the Swiss OFES and the European EQCSPOT project.
\end{acknowledgement}

\section{Figure Captions}

Figure 1

Experimental setup.

\bigskip

Figure 2

Gate pulse produced by the short gate generator.

\bigskip

Figure 3

Dark count probability per gate pulse of 2.4 ns versus detection efficiency.

\bigskip

Figure 4

Afterpulse probability per gate pulse of 2.4 ns versus time since previous
avalanche.

\bigskip

Figure 5

Afterpulse probability per gate pulse of 2.4 ns versus time since previous
avalanche.

\bigskip

Figure 6

Avalanche pulse of Epitaxx avalanche photodiode.

\bigskip

Figure 7

Afterpulse probability per 2 ns versus time since previous avalanche for
long gates (20 ns) of 3, 4, 6 and 8 V amplitude.

\bigskip

Figure 8

Dark count probability per 2 ns versus detection efficiency for long gates
(20 ns) of 3, 4, 6 and 8 V amplitude.

\bigskip

Figure 9

Timing jitter (full width at half maximum) versus detection efficiency.

\bigskip

Figure 10

QBER (solid lines) and normalized raw key creation rate (dashed line) versus
distance for Epitaxx APD at two temperatures, EG\&G and NEC APD's, and a
hypothetical photodiode with a dark count probability reduced by a factor of
ten ($2.8\cdot 10^{-6}$) with respect to that of the Epitaxx APD.

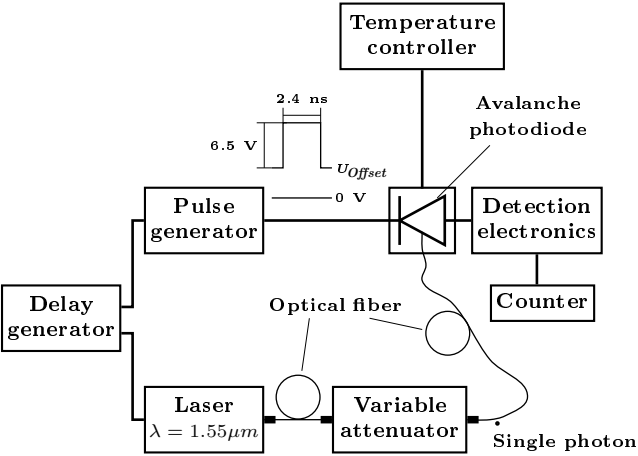
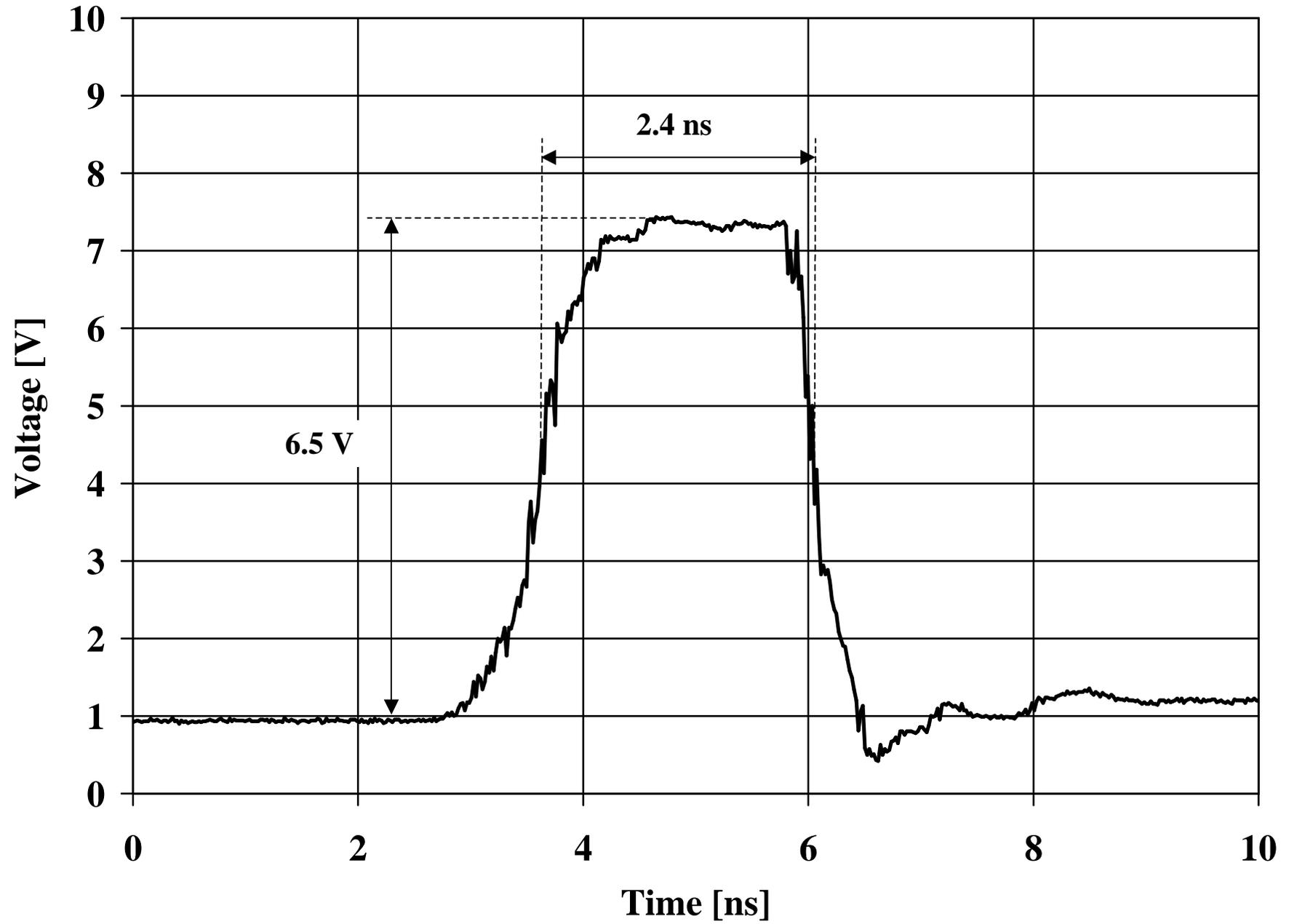
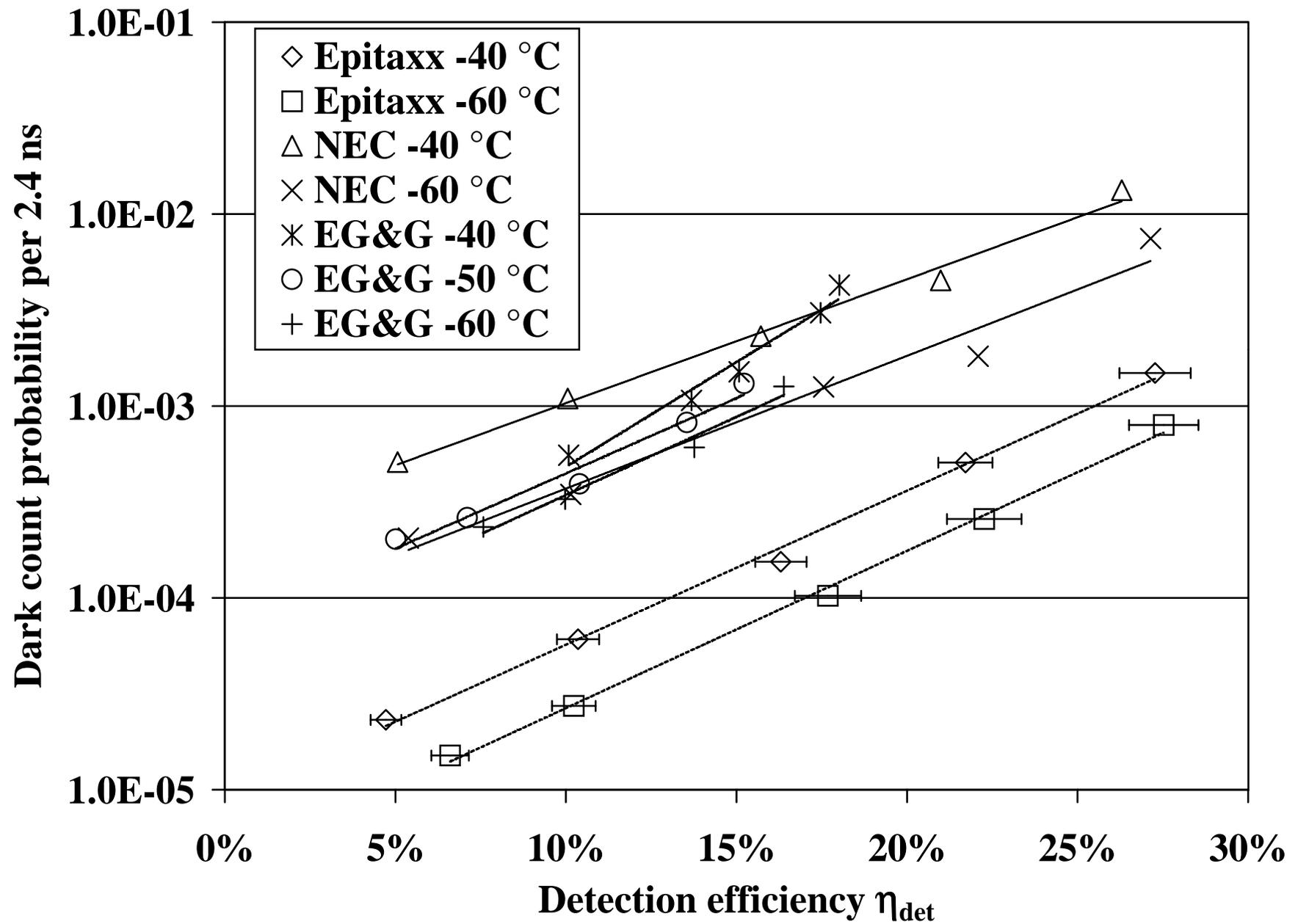
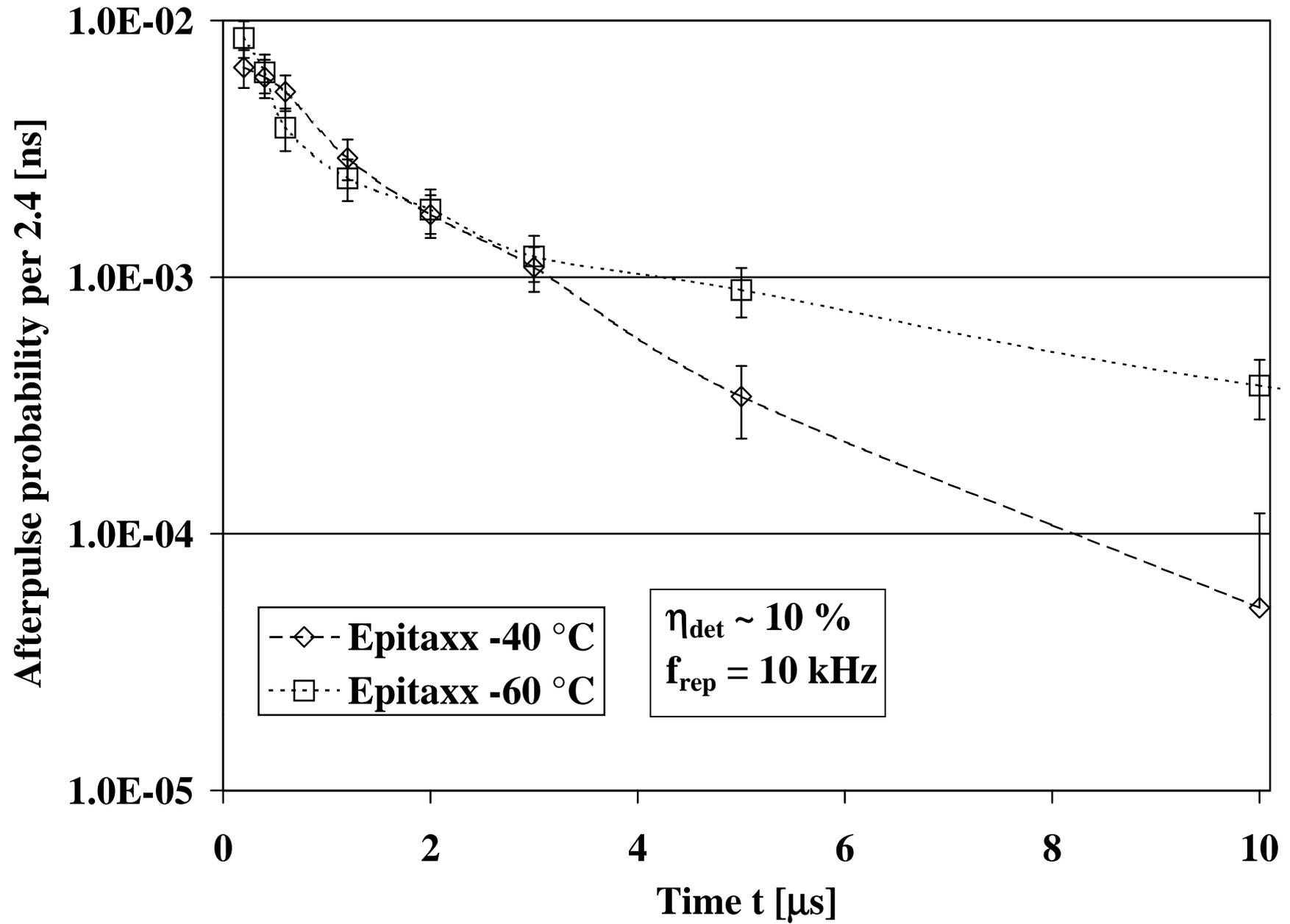
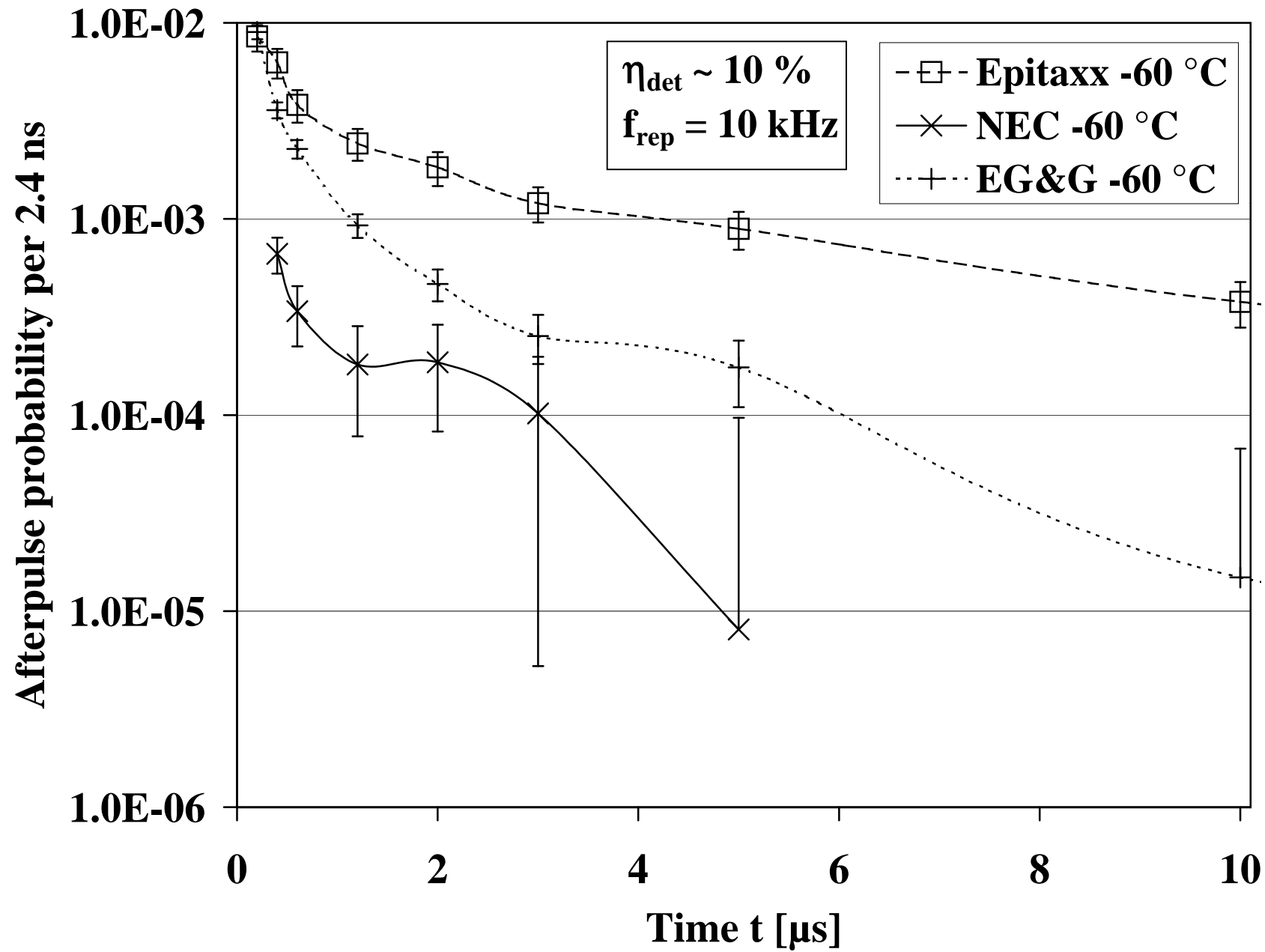
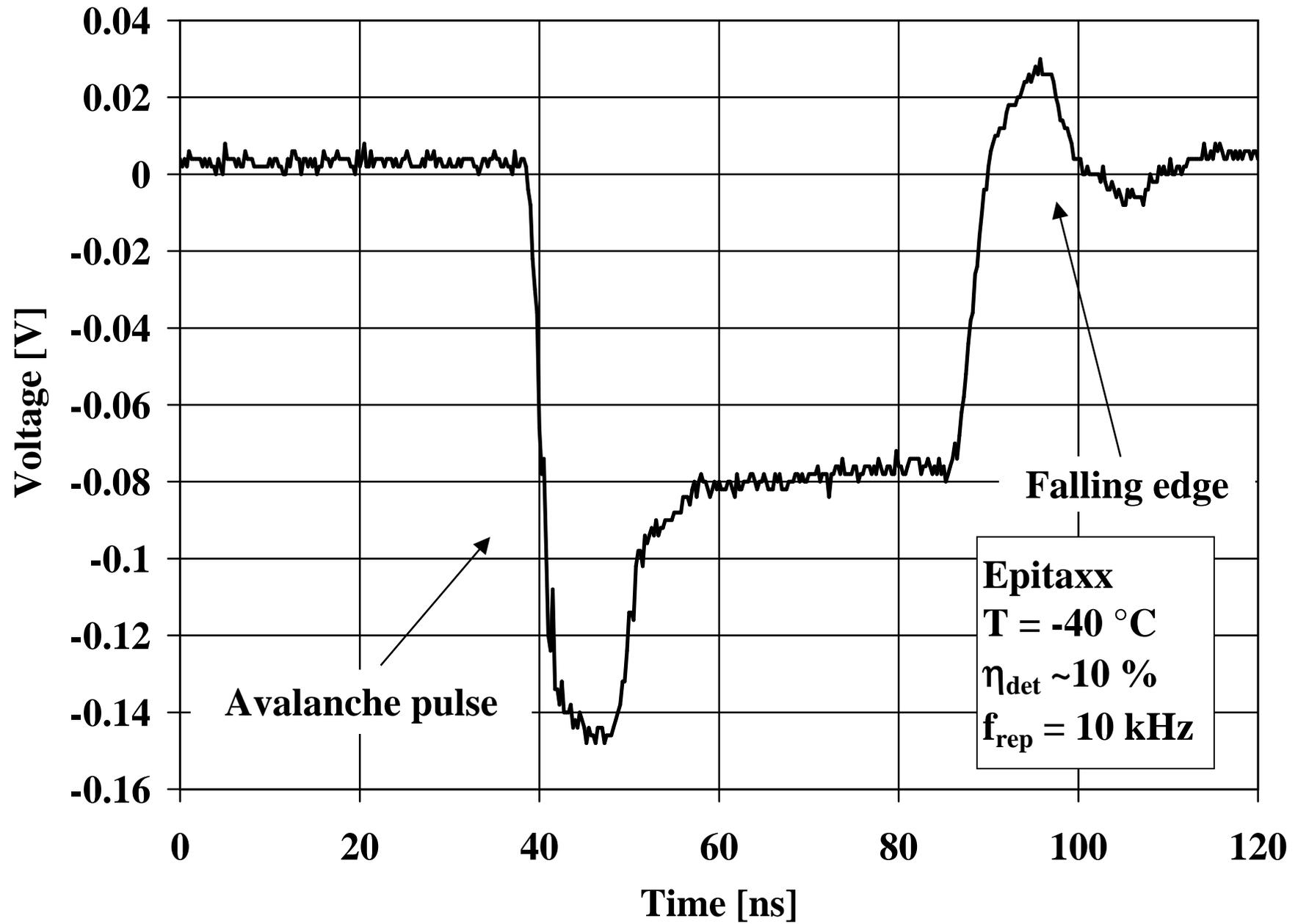
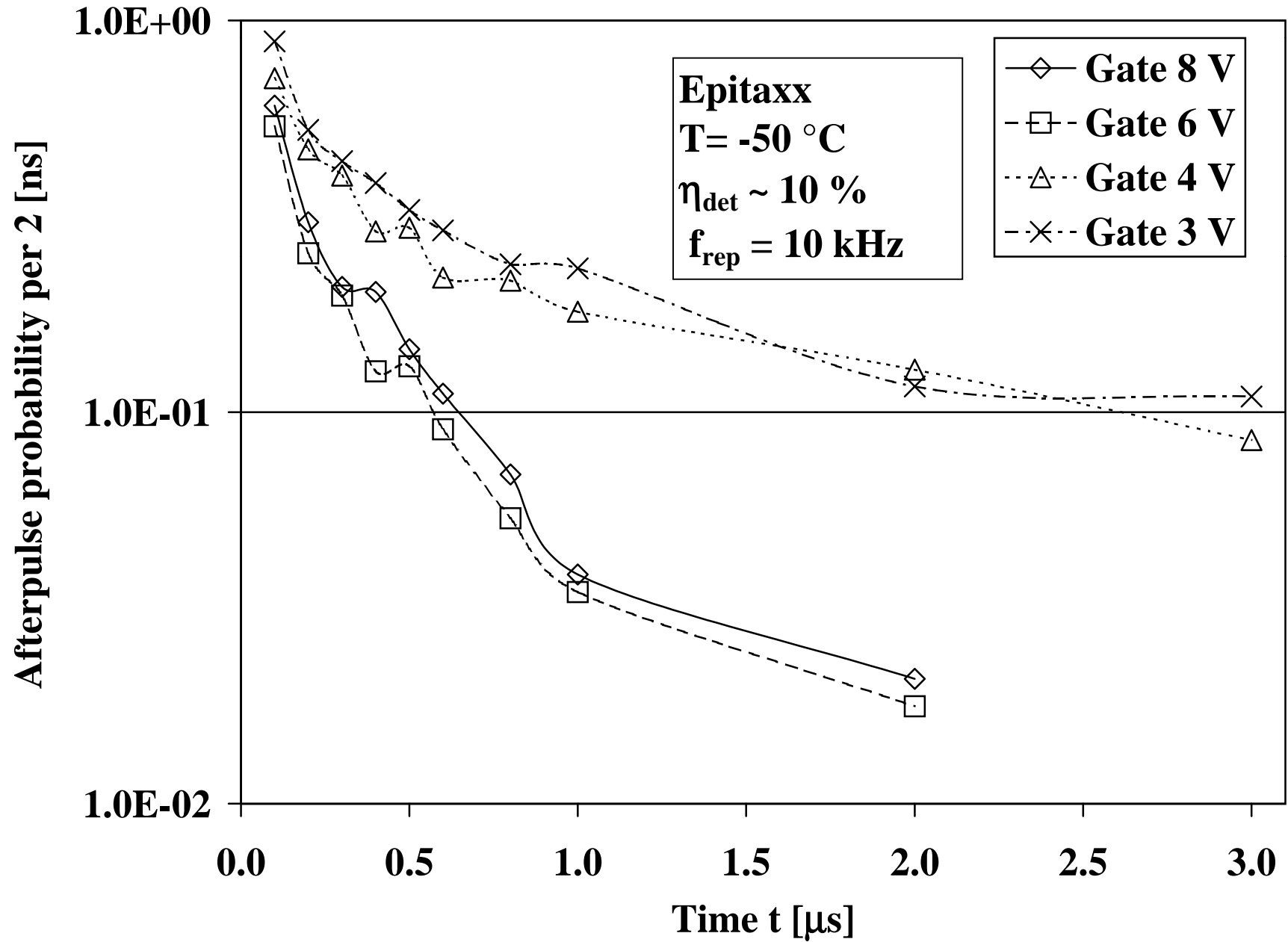
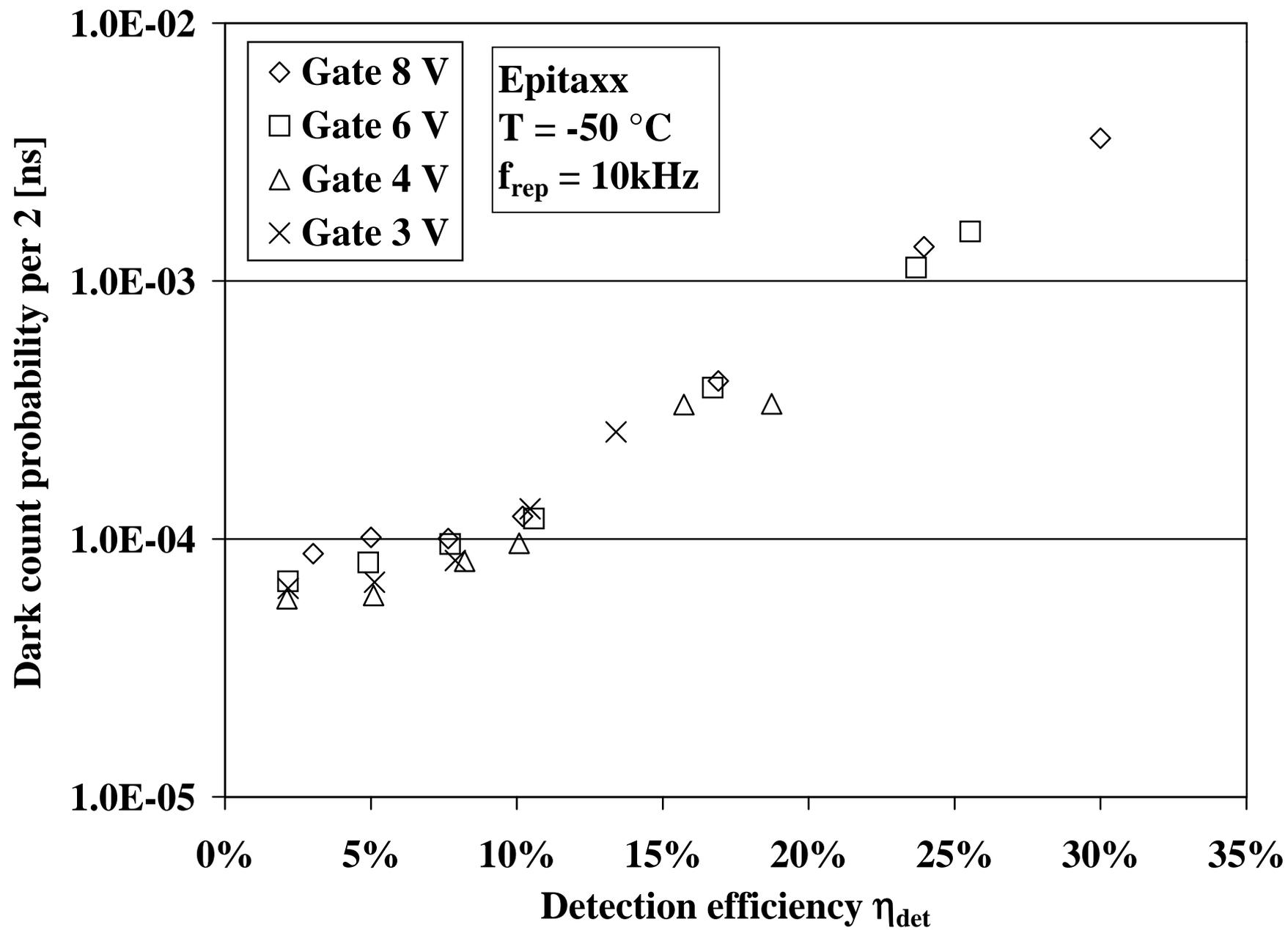
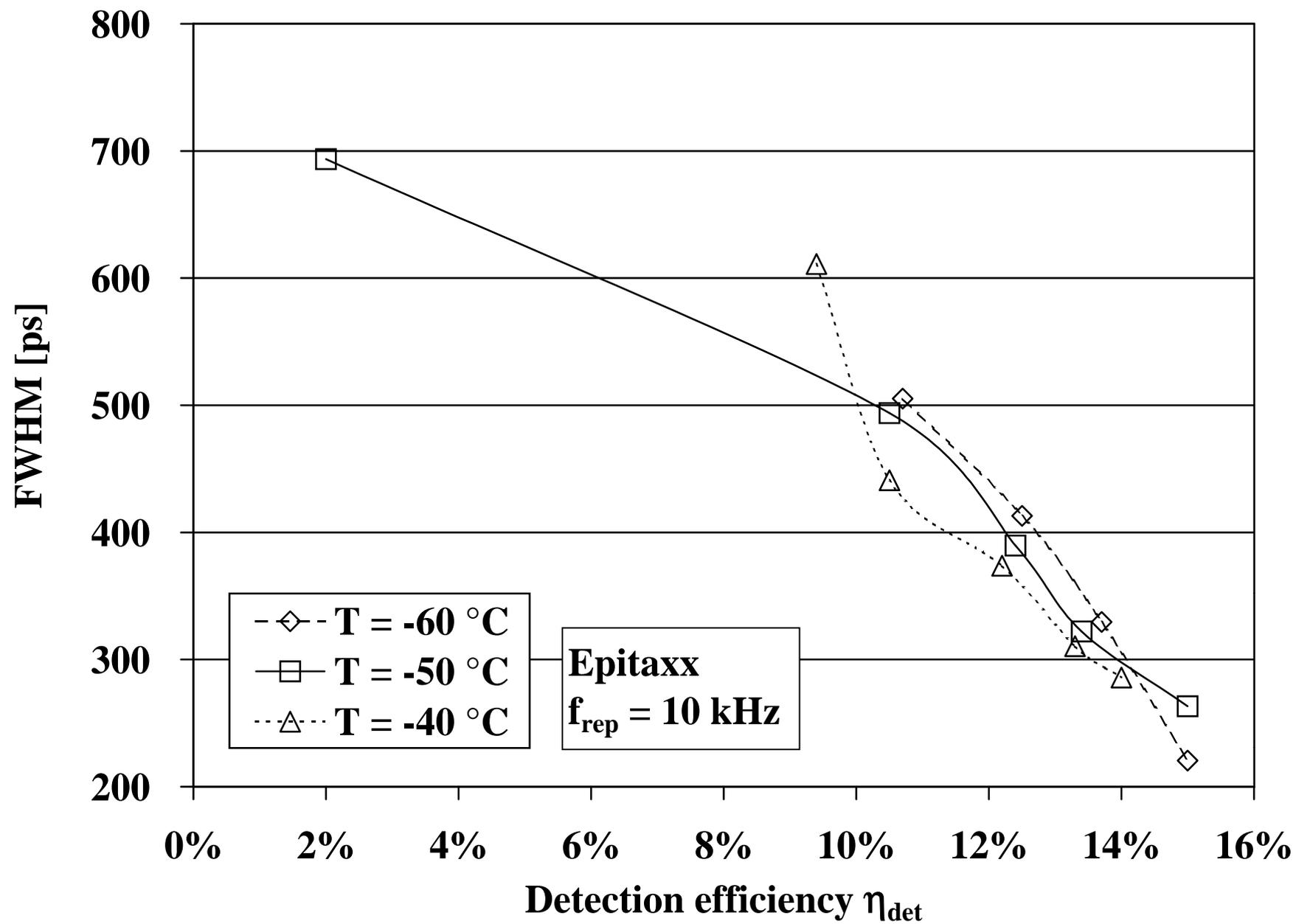
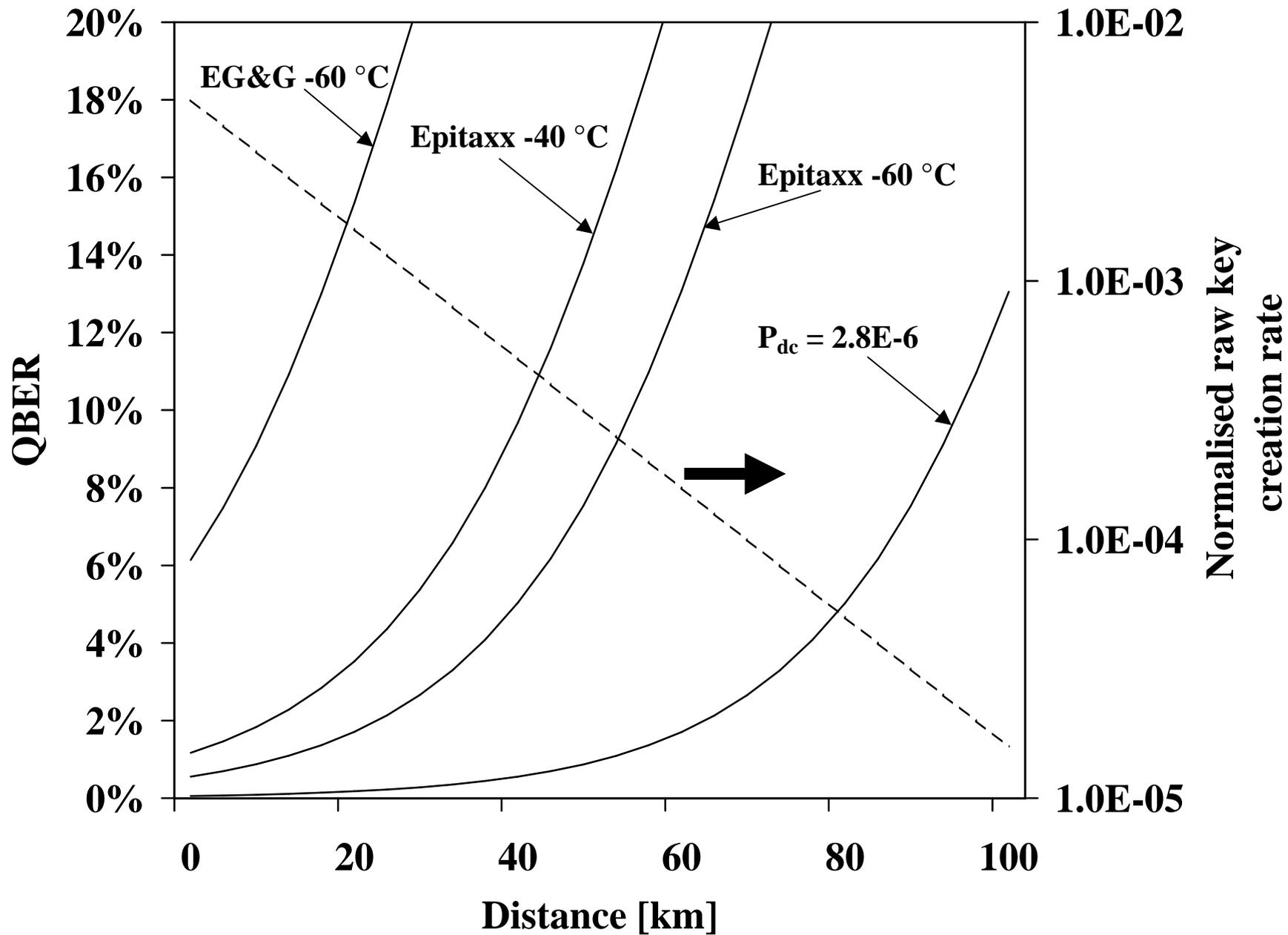
\end{document}